# Aspects of the Second Law of Thermodynamics from Quantum Statistical Mechanics to Quantum Information Theory

A. K. Rajagopal*, R. W. Rendell*, and Sumiyoshi Abe#

*Naval Research Laboratory, Washington, DC 20375, USA
#Institute of Physics, University of Tsukuba, Ibaraki 305-8571, Japan

**Abstract.** The Kullback – Leibler inequality is a way of comparing any two density matrices. A technique to set up the density matrix for a physical system is to use the maximum entropy principle, given the entropy as a functional of the density matrix, subject to known constraints. In conjunction with the master equation for the density matrix, these two ingredients allow us to formulate the second law of thermodynamics in its widest possible setting. Thus problems arising in both quantum statistical mechanics and quantum information can be handled. Aspects of thermodynamic concepts such as the Carnot cycle will be discussed. A model is examined to elucidate the role of entanglement in the Landauer erasure problem.

## I. INTRODUCTION

The laws of thermodynamics are concerned with the behavior of macroscopic objects. Deducing these laws from the quantum statistical mechanics of many particle systems requires certain identification of quantities that appear in the two formulations, as is well known [1]. The question of quantum limits to the thermodynamic laws is best answered by examining those problems where the quantum signature survives in the identification process alluded to above. In particular, the second law of thermodynamics is a statement about comparing two systems. In this paper, we employ the Kullback – Leibler inequality [2] (hereafter denoted by KL) which forms the basis for comparing any two density matrices. Stated in its general form,

$$K(\rho_A|\rho_B) \equiv Tr\rho_A(\ln\rho_A - \ln\rho_B) \geq 0, \qquad (1)$$

where Tr stands for the trace over the space over which the density matrices of two systems A and B are defined. In this expression, the two density matrices may, for example, be time-dependent or $\rho_A$ may be time-dependent being a solution of a quantum equation of motion and $\rho_B$, its stationary solution, or any other density matrix that may be represent the system. The classical version of the above inequality is stated in terms of two probability distributions where the trace is replaced by an appropriate integration (e.g., phase space or configuration space where the systems exist). In Quantum Information theory too, one often compares two situations as in the erasure or copying problems. From such a viewpoint, the second law in its widest possible context is a manifestation of the KL inequality.

A technique to set up the density matrix for studying the equilibrium properties of a physical system is to use the maximum entropy principle, given the entropy as a functional of the density matrix, subject to some known constraints. The time-dependent properties of the system are examined using a Master equation obeyed by the density matrix [3]. These three ingredients allow us to set up the laws of thermodynamics in its widest possible setting. Furthermore problems appearing in both quantum statistical mechanics and quantum information theory can be handled by these techniques. In the following, we will briefly examine four situations. In sec.II, we formulate the second law of thermodynamics by defining the density matrices in eq.(1) appropriately. We also introduce the definitions of quantity of heat, work, etc. in analogy with thermodynamics [1]. In this context, we will remark on the difference between the quantum and classical (-like) master equations as well. In sec.III, we compare two equilibria associated with two systems A and B, which in turn leads to ideas such as Carnot cycle, adiabatic and isothermal processes as in thermodynamics [3, 4]. In sec.IV, we examine a model of quantum information loss when bit is erased from a quantum entangled qubit system (Landauer's problem) [5]. In sec.V, we consider an entropy functional (non-additive) different from the usual von Neumann type [6] to point out how general these considerations are.

## II. DENSITY MATRIX, MASTER EQUATION, DEFINITIONS

The most general quantum master equation which is linear, local in time, preserves the essential properties of the density matrix, namely hermiticity, traceclass (probability conservation) structure, and positive semi-definiteness is the Lindblad equation [3] (units where the Planck constant, $\hbar$ is set equal to 1):

$$\partial\rho/\partial t = -i[H,\rho] + \frac{1}{2}\sum_j k_j \left\{ 2L_j\rho L_j^+ - L_j^+L_j\rho - \rho L_j^+L_j \right\} \qquad (2)$$

where the first term in the right hand side represents the unitary time evolution driven by H, a hermitian Hamiltonian operator and the second term represents dissipative evolution if all the real parameters $k_j$ are positive, accomplished by the operators $L_j$ along with their hermitian conjugates, $L_j^+$. This choice of operators is not unique and the sum over j is as yet unspecified. The unitary time evolution part may be subsumed by a transformation $\rho_d = U\rho U^+$, so that the new equation has no H-term but has modified L-operators, $\tilde{L}$. This equation has two additional features, one, it can take a pure state into a mixed state and vice versa, and two, it preserves positivity of the density matrix throughout the evolution. This last property is often violated in phenomenological master equations. Since the density matrix is hermitian, a diagonal representation for it can be chosen $\rho_d = \sum_\alpha |\alpha\rangle p(\alpha)\langle\alpha|$, where $\{|\alpha\rangle\}$ is an orthonormal set for each instant of time. $p(\alpha)$ have the same physical interpretation as the probabilities in a classical description of the system.

With the choice of the entropy functional in the von Neumann form
$$S_1[\rho] = -Tr\rho \ln \rho = -\sum_\alpha p(\alpha) \ln p(\alpha) \tag{3}$$

we now calculate $\partial S_1/\partial t$:
$$\partial S_1/\partial t = -Tr\partial \rho/\partial t \ln \rho = \sum_{\alpha,\alpha'} K_{\alpha\alpha'} p(\alpha')(\ln p(\alpha') - \ln p(\alpha)) \geq 0 \tag{4}$$

provided the Lindblad operators are hermitian. Here $K_{\alpha\alpha'} = \sum_i k_i |\langle\alpha|\tilde{L}_i|\alpha'\rangle|^2$. This is the KL relation, eq.(1), and follows after using $\ln x \geq 1 - 1/x, x > 0$, and upon interchanging summation indices and definitions. Thus the increase of entropy is guaranteed under the above condition. Under such a condition, one also finds that the right hand side of modified eq.(2) takes the form, $-\frac{1}{2}\sum_j k_j [\tilde{L}_j, [\tilde{L}_j, \rho_d]]$.

Another important aspect of eq.(2) concerns its stationary solution, $\rho_s$, given by $\partial \rho/\partial t = 0$. This is also the asymptotic, long time solution of eq.(2). A particular form of the stationary solution obeys the "detailed balance" relation given by
$$K_{ab;s} p_s(b) = K_{ba;s} p_s(a) \text{ for all pairs } a,b. \tag{5}$$
The subscript s here denotes the stationary limit of the matrix elements introduced here:
$$\partial \rho_{ds}/\partial t = 0 = \frac{1}{2}\sum_j k_j \{2\tilde{L}_j \rho_{ds} \tilde{L}_j^+ - \tilde{L}_j^+ \tilde{L}_j \rho_{ds} - \rho_{ds} \tilde{L}_j^+ \tilde{L}_j\}. \tag{6}$$
The stationary density matrix can also be expressed in terms of its diagonal representation as before: $\rho_{ds} = \sum_a |a\rangle p_s(a) \langle a|$, $\{|a\rangle\}$ an orthonormal set for asymptotic, stationary time limit. These are in general different from those defined before at any given instant, t. These states now appear in defining the detailed balance condition in eq.(5) above.

We may compare the general solution with the asymptotic solution by considering the KL relation $K(\rho|\rho_s) \equiv Tr\rho(\ln \rho - \ln \rho_s) \geq 0$. The time derivative of this expression is not easy to manipulate because of the two different representations of the density matrices. We obtain $\partial K(\rho|\rho_s)/\partial t \leq 0$, after making some technical assumptions [4, 7]. Thus the general solution approaches the asymptotic one if the detailed balance holds.

The mean value of any physical quantity, for example the system energy, is defined by $U = Tr\rho H = \sum_\alpha \langle\alpha|H|\alpha\rangle p(\alpha)$. If we employ the maximum entropy principle subject to this given mean energy, then the resulting density matrix is given by
$$\rho_{eq} = \exp-(\beta H)/Z, \ Z = Tr\exp-\beta H \tag{7}$$
where $\beta$ is a Lagrange multiplier, the inverse temperature in units where the Boltzmann constant is taken to be unity. Z here is the partition function. In this theory then, the equilibrium density matrix given in eq.(7) leads to the known thermodynamic relation between the free energy, F, the entropy, S, and the mean energy, U:
$$F = -\beta^{-1} \ln Z = U - \beta^{-1} S, \quad S \equiv S_1[\rho_{eq}].$$

A general variation of this average value can come about by variations in H and in $\rho$, $dU = Tr(d\rho)H + Tr\rho(dH)$. Following [1], we may identify "work", $dW = Tr\rho(dH)$, and "quantity of heat", $dQ = Tr(d\rho)H$, quite generally. If we consider the variation to come about by time evolution, then we find that

$$dS_1/dt - \beta \, dQ/dt = -\sum_\alpha \frac{dp(\alpha)}{dt}\left(\ln p(\alpha) - \ln p_{eq}(\alpha)\right) = -\frac{dK(\rho|\rho_{eq})}{dt} \geq 0 \qquad (8)$$

where we used the KL relation between the density matrix and its equilibrium counterpart, eq.(7), and the result mentioned above. This type of thermodynamic result was deduced in [4] for weak coupling of the system with the heat bath. In the next section, we apply similar methods to derive various results of thermodynamics.

## III. QUASI-STATIC PROCESSES, CARNOT CYCLE etc.

By considering two equilibrium density matrices of the form given by eq.(7) described by its internal energies specified by two different Hamiltonians and temperatures, in the KL eq.(1), we deduce the quasi-static thermodynamic relations. These lead to Carnot cycles etc. The associated free energies and entropies of the two systems are given correspondingly. Then, we obtain

$$K(\rho_1|\rho_2) = -\beta_1 Tr(\rho_1 H_1) + \beta_2 Tr(\rho_1 H_2) + \ln Z_2 - \ln Z_1 \geq 0. \qquad (9)$$

This inequality in terms of entropies is then

$$K(\rho_1|\rho_2) = S_2 - S_1 + \beta_2 Tr(\rho_1 - \rho_2)H_2 \geq 0. \qquad (10)$$

Defining the quantity of heat in going from system 1 to system 2 by recalling such a definition from above, $\Delta Q_{21} \equiv Tr(\rho_2 - \rho_1)H_2$, with the understanding that system 2 has a higher entropy than system 1, we have an interesting result

$$\Delta Q_{21} \leq \beta_2^{-1}(S_2 - S_1). \qquad (11)$$

It should be noted that this result is equivalent to a variant of eq.(8).
On the other hand, expressing eq.(9) in terms of free energies, we obtain another inequality in terms of work done by recalling such a definition from before, by considering an isothermal process when the temperatures of the two systems are the same, $\Delta W_{21} \equiv Tr(H_2 - H_1)\rho_2$:

$$\Delta W_{21} \leq F_2 - F_1. \qquad (12)$$

Let us consider an adiabatic process consisting of a return path 3 to 4 at a lower temperature such that the entropy at 3 is the same as that at 2 and entropy at 4 is the same as at 1 (isoentropic). Then eq.(11) represents heat supplied to the system while the return path gets heat out of the system. The calculation proceeds in the same manner as above and the result is $\Delta Q_{43} \leq \beta_1^{-1}(S_2 - S_1)$. The Carnot efficiency is then defined as the ratio of the quantity of heat in minus the heat out divided by the input heat, which is the standard result found by thermodynamic methods. In the next section, we consider the problem of quantum information loss by applying the above techniques to a model quantum entangled system.

## IV. THE LANDAUER ERASURE PROBLEM: A MODEL OF AN ENTANGLED QUBIT

We now turn our attention to the Landauer erasure problem [8, 9, 10], that of bounds on the heat generated when information (in units of classical or quantum bits) is erased, and consider this problem when the information is encoded in a quantum entangled way [5].

Consider an isothermal erasure process: the system is in thermal equilibrium initially and it is forced into a known state by tuning parameters in the system Hamiltonian. Initially we assume the system is in a state given by a density matrix, $\rho_i$, and its entropy is $S(\rho_i)$. Upon tuning the parameters the density matrix becomes $\rho_f$. Thus the erasure process leads to a change in entropy of the whole system given by

$$\Delta S = S(\rho_f) - S(\rho_i). \tag{13}$$

A model of two coupled qubits is generated by the Hamiltonian $H = (\Delta\varepsilon_1 \sigma_{1z} \otimes I + \Delta\varepsilon_2 I \otimes \sigma_{2z})/2 + J(\sigma_{1+}\sigma_{2-} + \sigma_{1-}\sigma_{2+})$, $\Delta\varepsilon_i$ is the energy level spacing of qubit i and J is the coupling strength and the qubits are represented by Pauli spin matrices for spins 1 and 2. The two-qubit density matrix and the concurrence, C, can be calculated at temperature T. The concurrence is a measure of entanglement of the two qubits and is a function of the Hamiltonian parameters[11]. C=0 for J=0 and denotes unentangled state whereas C≤1 for finite J but its value also depends on other parameters. We now examine the conditions under which qubit 1 can be erased. For this purpose we will take qubit 1 initially in a state with $\beta\Delta\varepsilon_{1i} \ll 1$, and force it into its ground state by increasing $\Delta\varepsilon_1$ so that its final state has $\beta\Delta\varepsilon_{1f} \gg 1$. Then the entropy change for the erasure is the entropy difference associated with the two qubit density matrices thus constructed. When J=0, we have two uncoupled qubits and we recover the Landauer bound of ln2 for the entropy difference. For J>0, entangled erasure is the distentanglement of an initally entangled state into an effectively unentangled erased state and bounds on erasure and disentanglement are found to be related. The cost of erasure is due to work done against entanglement.

## V. CONSIDERATIONS OF OTHER ENTROPY FUNCTIONALS

Recently there has been a surge of interest in examining the statistical mechanics of systems which do not admit of Gibbsian, exponential class of probability distributions, but are power-law distributions. Systems requiring such descriptions are fractals, anomalous diffusion processes, etc. They may also be relevant to nanosystems where nonadditivity of physical properties due to surface dominating features for instance, need to be recognized. The reader may refer to [6] for a description of this framework. In such situations, it is found that the ordinary probabilities need to be replaced by "escort" probabilities, defined by $P_q = \rho^q / Tr\rho^q$. The considerations of thermodynamic-like description of these systems is possible by following steps similar to those given in sec. II, for defining ideas such as quantity of heat etc. The results so deduced are in agreement with those given in [12].

As is well known [6], the maximum entropy method given in sec.II is applicable to other forms of entropy functionals with constraints defined differently in terms of the escort probabilities. In particular, in the non-additive (non-extensive) context, we employ the Tsallis form for the entropy with an index q: $S_q = [Tr(\rho)^q - 1]/(1-q)$, which goes to the von Neumann form, eq.(3) when q=1. The result corresponding to eq.(7) then is

$$\rho_q = [1 - (1-q)\tilde{\beta}(H - U_q)]^{1/(1-q)} / Z_q, \quad \tilde{\beta} = \beta/c_q, \quad c_q = Tr\rho^q, \tag{14}$$

with $U_q = TrP_q H$. Following the same steps as in sec.II, we deduce $dS_q = \tilde{\beta} dQ_q, \quad dQ_q = dU_q - TrP_q dH$, which was also obtained in [12].

We may conclude the paper with the remark that the KL inequality, eq.(1), has been used here to provide a unifying framework to discuss inequalities that play central roles in both the statistical thermodynamic and information theoretic contexts.

## ACKNOWLEDGMENTS

The first two authors are supported in part by the US Office of Naval Research. It is a pleasure to acknowledge that the participation of AKR in this Conference was possible because of a generous scholarship from the Conference.

## REFERENCES

1. F. Reif, *Statistical and Thermal Physics,* Sec. 6.6, (McGraw-Hill, New York, 1965).
2. S. Kullback and R. A. Leibler, Ann. Math. Stat. **22**, 79 (1951); S. Kullback, *Information Theory and Statistics* (Wiley, New York, 1959).
3. G. Lindblad, Comm. Math. Phys. **48**, 119 (1976); V. Gorini, A. Kassowski, and E. C. G. Sudarshan, J. Math. Phys. **17**, 821 (1976). See T. Banks, L. Susskind, and M. E. Peskin, Nucl. Phys. **B244**, 125 (1984).
4. R. Alicki, Lecture at 38 Winter School of Theoretical Physics. Ladek, Poland, e-print: quant-ph/0205188 (31 May 2002). See also A. K. Rajagopal and R. W. Rendell, Physica, **A305**, 297 (2002).
5. R. W. Rendell and A. K. Rajagopal, in preparation (2002).
6. Several articles in S. Abe and Y. Okamoto (Eds.) *Nonextensive Statistical Mechanics and Its Applications* (Springer, New York, 2001).
7. A. K. Rajagopal, Phys. Letts. **A246**, 237 (1998).
8. V. Vedral, Proc. Roy. Soc. (London) **A456**, 969 (2000).
9. K. Shizume, Phys. Rev. **E52**, 3495 (1995).
10. B. Piechocinska, Phys. Rev. **A61,** 062314 (2000).
11. X. Wang, Phys. Rev. **A64**, 012313 (2001).
12. S. Abe, Physica **A300**, 417 (2001); Ibid, **A305**, 62 (2002).